# Advanced Ocean Reanalysis of the Northwestern Atlantic: 1993-2022

Ruoying He[1], Tianning Wu[1], Shun Mao[1], Haibo Zong[2], Joseph Zambon[2], Jennifer Warrillow[2], Jennifer Dorton[3], Debra Hernandez[3]

1. Department of Marine, Earth and Atmospheric Sciences, North Carolina State University
2. Fathom Science, Inc.
3. Southeast Coastal Ocean Observing Regional Association

Corresponding author: Ruoying He (rhe@ncsu.edu)

## Abstract

A 30-year high-resolution Northwestern Atlantic Ocean Reanalysis (NAOR) is presented. NAOR spans from January 1993 to December 2022 with a 4 km horizontal resolution and 50 vertical layers. It provides enhanced resolution and expands the spatial and temporal coverage of existing ocean reanalysis in the region. NAOR was conducted using the Regional Ocean Modeling System along with Ensemble Optimal Interpolation data assimilation. Open boundary and surface forcing conditions were obtained from GLORYS global ocean reanalysis and ECMWF ERA5 reanalysis. Multiple sources of satellite and in-situ observations were incorporated through the data assimilation. Additionally, major rivers were accounted for to include freshwater riverine discharge. NAOR was extensively evaluated against available independent observations. Spatio-temporal variations of mesoscale circulation, eddies, and boundary currents are well captured. Compared to GLORYS, NAOR offers a more accurate physical and dynamic baseline of the northwestern Atlantic Ocean, which can be utilized for a range of marine and environmental studies as well as climate impact research.

## Background and Summary

Ocean reanalysis (OR) has emerged as a crucial tool for elucidating ocean dynamics and assessing climate change impacts. Recent years have seen an increase in coastal OR studies, which assimilate historical observations into numerical models to synthesize coastal and deep ocean data, resulting in time-varying, three-dimensional reconstructions of ocean conditions. By reconstructing past ocean states, detailed comparisons can be made of changes and variations in ocean processes over a wide range of time scales.

In the United States, two notable OR studies have been conducted. Ref [1] performed the California Current System State Estimate, encompassing a 32-year period from 1980 to 2012. This study incorporated satellite data, Argo floats, and in-situ measurements to provide a comprehensive analysis of variability and trends in ocean temperature, salinity, and circulation along the California coast. Ref [2] conducted the Northeast U.S. Coastal OR, spanning from 2007 to 2021, offering detailed insights into historical ocean conditions in the northeast U.S. coastal region, including coastal sea level and bottom temperature variability.

Australian researchers also have contributed to coastal OR research. Ref [3] focused on an 18-year Bluelink ReANalysis (BRAN) in the Australian region, highlighting climate change impacts on ocean temperature, salinity, and circulation in this system. Ref [4] conducted a reanalysis of the Great Barrier Reef and Coral Sea from 2006 to 2015. Their work illustrates seasonal and climatic processes driving variability in ocean conditions such as El Niño and La Niña and their influence on sea temperature and freshwater flux to the coastal environment.

European coastal OR studies have been equally significant. For example, Ref [5] produced a 30-year reanalysis of the Baltic Sea from 1970 to 1999, improving the estimation of halocline and thermocline depths. Ref [6] conducted a Mediterranean reanalysis from 1987 to 2019, examining variability in Mediterranean Sea conditions, including the effects of climate patterns on regional ocean heat and salt contents. The Copernicus Marine Environment Monitoring Service (CMEMS) offers multiple long-term regional reanalysis products, including the North-West European Shelf, Baltic Sea, and Black Sea, facilitating comprehensive understanding of changes in these regional water bodies[7]. These dedicated regional reanalysis efforts offer improved ocean state representation over global reanalysis products, providing critical values for understanding and predicting regional scale ocean dynamics and marine biogeochemical phenomena.

These and other studies demonstrate the efficacy of OR in investigating ocean dynamics and climate change effects on the coastal ocean globally. Building on this foundation, the present study introduces a new 30-year OR of the northwestern Atlantic Ocean (Figure 1), spanning from January 1993 to December 2022. A 4-km resolution provides a significant improvement in capturing fine-scale ocean dynamics, such as tides, river inputs, detailed coastlines, and complex bathymetry, which are crucial for accurate regional oceanographic studies. This work aims to provide a realistic physical and dynamic baseline for the northwestern Atlantic marginal seas, including the North American east coast shelf seas, the Gulf of Mexico, and the Caribbean Sea. It fills the gap between the demand for high-resolution and accurate ocean dynamics in coastal ocean studies and coarser-resolution global reanalysis, serving as a valuable resource for regional marine and environmental studies as well as climate impact research.

## Methods

### Model

The Northwestern Atlantic Ocean Reanalysis (NAOR) is based on the Regional Ocean Modeling System (ROMS) [8] The model is configured on a rectangular grid encompassing the northwestern Atlantic Ocean (Figure 1). The model employs a horizontal resolution of 1/25th degree (approximately 4 km) and 50 vertical sigma levels, with enhanced resolution near the surface and bottom to better resolve boundary layer dynamics. This configuration is designed to adequately capture mesoscale circulation, eddies, and boundary currents within the study area. Model bathymetry was constructed from the General Bathymetric Chart of the Oceans (GEBCO) 15 arc-second global bathymetric grid. Despite the high-resolution model grid, the criterion for hydrostatic primitive equation remains satisfied in our model. As described in ref [9], hydrostatic primitive equation is valid under the condition that the aspect ratio between vertical and horizontal length scales is less than one. In NAOR, the vertical length scale is approximately 1000 meters, and the horizontal length scale is approximately 4000 meters. Thus, for our purpose of capturing mesoscale ocean circulation features, the use of hydrostatic ROMS is appropriate and supported by peer-reviewed publications[10].

NAOR used five primary inputs. The meteorological forcing was obtained from European Centre for Medium-Range Weather Forecasts (ECMWF) Reanalysis v5 (ERA5) [11]. CMEMS Global Ocean Physics Reanalysis (GLORYS) 1/12th degree daily output [12] provided both the initial conditions on January 1, 1993, and open boundary conditions through December 31, 2022. Freshwater influx from 120 major rivers was obtained from the National Water Model for U.S. rivers and historical climatologies for rivers outside the U.S. Tidal forcing for sea surface height and barotropic velocities was derived from the Oregon State University TOPEX/POSEIDON global tidal model (TPXO), incorporating 10 major tidal constituents [13].

Hourly ERA5 atmospheric fluxes were incorporated using the COARE 3.0 bulk flux algorithm [14]. At the open boundary, a combined nudging-radiation condition was implemented for temperature (T), salinity (S), and velocity components (U and V). The Chapman condition [15] was applied for the free-surface ($\eta$), while the Flather condition was used for 2D momentum flux [16]. Horizontal advection of momentum utilized a third-order, upstream biased scheme [17], whereas horizontal and vertical tracer advection used the multidimensional positive definite advection transport algorithm (MPDATA) [18]. The Smagorinsky scheme [19] was selected for spatially dependent horizontal mixing of momentum and tracers, and the Large-McWilliams-Doney (LMD) scheme [20] was selected for vertical mixing, primarily for stability and computational efficiency. Bottom drag was parameterized using the quadratic law. The minimum depth was set to 10 m, and the wetting/drying conversion scheme was disabled.

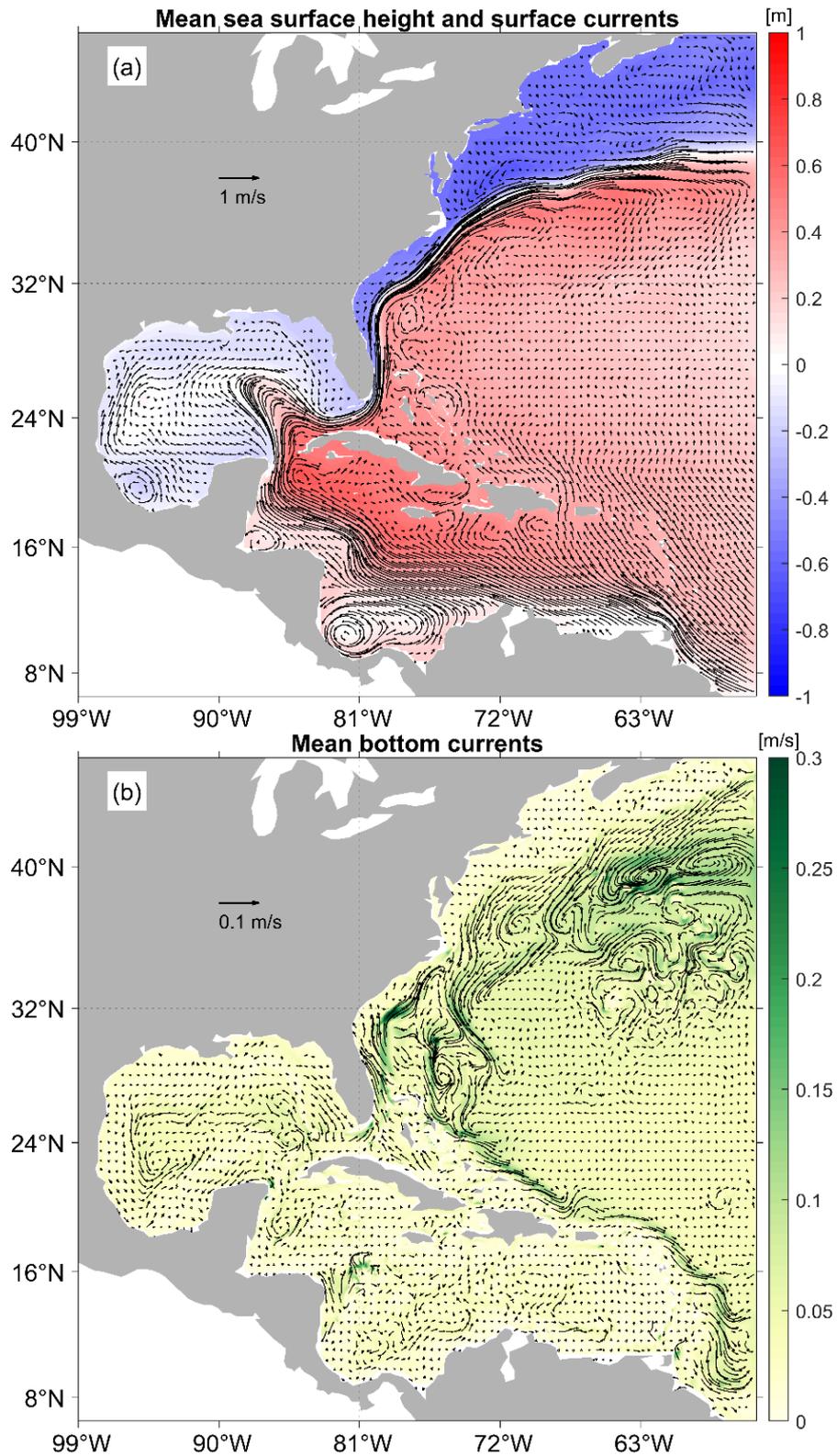

Figure 1. The Northwestern Atlantic Ocean Reanalysis (NAOR) domain. (a) 30-year (1993-2022) mean sea surface height (SSH; color shading) and mean surface current vectors (black arrows). (b) Mean bottom current speed (color shading) and bottom current vectors (black arrows).

**Observations**

The assimilated observations in this study comprise altimetry Sea Level Anomaly (SLA), remotely sensed Sea Surface Temperature (SST), and in-situ temperature and salinity (T/S) profiles from shipboard conductivity-temperature-depth (CTD) sensors, expendable bathythermograph (XBT), autonomous underwater vehicles (AUVs), and Argo floats (Figure 2). SLA data were obtained from Archiving, Validation, and Interpretation of Satellite Oceanographic data (AVISO) and subjected to ocean tide and mean dynamic topography corrections. SST data were obtained from AVHRR Pathfinder Version 5.3 L3C Daily 4 km SST and MODIS Terra L3 Daily 4 km SST. T/S profiles were sourced from the UK Met Office EN4 quality controlled subsurface profiles and the Integrated Ocean Observing System (IOOS) AUV dataset. The latter provides additional detailed nearshore observations particularly in North American east coast and Gulf of Mexico.

A super-observation procedure was implemented for each variable[21]. All observations within a model grid cell, irrespective of instrument or platform, were combined and weighted by the inverse observation error variance. These weighted observations were then averaged to obtain a representative value for the location and time within the cell and observation window.

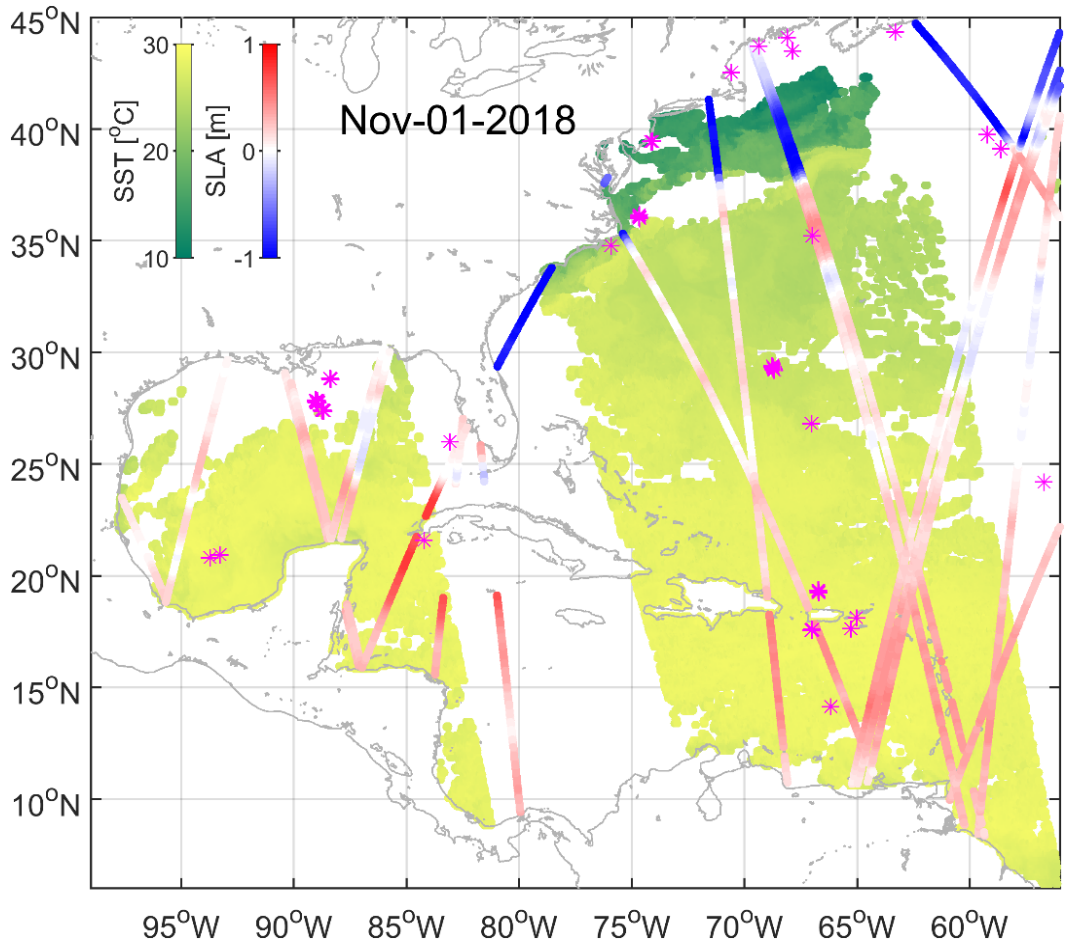

Figure 2. Spatial distribution of observations on November 1, 2018. Super-observations, representing time and space-averaged composites at model horizontal grid resolution, are depicted. The color shading from yellow to green indicates SST. Red and blue lines represent altimetric SLA. Asterisks denote the locations of T/S profiles collected by ship surveys, AUVs, and Argo floats on this specified date.

**Data Assimilation**

Data assimilation (DA) is a method to incorporate observational information into models, with the objective of improving forecast accuracy by making the modeled state more consistent with observations. Two main approaches are used for DA: the 4-Dimensional Variational Data Assimilation (4DVAR) method and the Ensemble Data Assimilation (EnDA) method.

The 4DVAR method corrects the model's initial conditions, boundary conditions, and surface forcing while preserving the system's dynamical balance, allowing the resulting fields to be fed into the same analysis of term balances as a free-run model [22]. Conversely, the EnDA method constrains the model towards a state more consistent with the observations [23]. Owing to the ensemble, EnDA handles model uncertainties and errors very well, especially in highly nonlinear systems. It is also less computationally expensive for large-scale systems.

We have successfully employed the 4DVAR method in a suite of regional OR studies [24–27]. However, this method is computationally intensive and time-consuming due to its requirement for iterative integration between forward and adjoint models. For long-term reanalysis, EnDA provides a more efficient approach for operational data assimilative forecasting by efficiently incorporating observed data and constraining the model towards a state more consistent with the data.

In NAOR, the ocean state vector is represented as x = x(η, T, S, U, V, ū, v̄), where ū and v̄ denote the components of the 2D depth-averaged velocity. Initially, a five-year hindcast (control run) was performed to provide an initial estimate of oceanographic conditions and to compute background error covariances for the DA process. This hindcast inherits potential skill from the global ocean reanalysis and atmospheric reanalysis used in forcing the OR.

The subsequent reanalysis consists of a series of short-term forecasts aimed at achieving a dynamic balance between model and observations through data assimilation and initialization steps. The analysis equation (Eqn. 1) and background covariances (Eqn. 2) for the ensemble-based DA are:

$$\mathbf{x}^a = \mathbf{x}^f + \mathbf{B}\mathbf{H}^T[\mathbf{H}\mathbf{B}\mathbf{H}^T + \mathbf{R}]^{-1}[\mathbf{y} - \mathrm{L}(\mathbf{x}^f)] \tag{1}$$

$$\mathbf{B} \equiv \mathbf{A}\mathbf{A}^T[(\mathbf{m} - 1)]^{-1} \tag{2}$$

where $x^a$ and $x^f$ denote the analysis and forecast vectors respectively, and the vector $y$ represents the observation. Other symbols include linearized observation operator **H**, background error covariance **B**, observation error covariance **R**, ensemble anomalies **A**, and ensemble size **m**. To generate ensemble anomalies, 155 ensemble members were constructed from multi-year hindcast data, each spaced 7 days apart, using a 3-day running mean minus a 30-day running mean. This approach effectively filters out dependencies ranging from sub-daily to multi-week, representing background error covariances at appropriate spatio-temporal scales for the assimilated observations and the 3-day analysis-forecast cycle. Forecast innovations were calculated using daily forecast fields at observation time.

Sensitivity experiments were conducted to fine-tune variable-dependent localization radii, R-factors, and K-factors. The K-factor was employed to adaptively limit the increment to ensemble spread, while the R-factor, defined as a multiple of observation error variance, was used to tune the Kalman gain to achieve a more balanced observation error-ensemble spread relationship. As suggested by ref [28], the K-factor also serves to avoid large increments in the analysis and improve balance without discarding associated observations that could provide important information to the system.

In the DA procedure, the initial condition is updated to allow the model to run the complete length of each DA cycle as a dynamical forecast. The increment, defined as the difference between $x^f$ and observations, is then added to the ocean state, forming the initial condition for the next analysis cycle. To ensure that all calculated errors and increments are based on model dynamics, each forecast is run for an additional 1.5 days past the base date of the next analysis. The centered observation window is compared directly with the corresponding observations.

Data vendors provide observation error estimates (R) for SLA and SST, which are 0.02 m and 0.5 °C, respectively. The observation error estimates for in-situ T/S profiles are fixed at 0.5 °C and 0.1 psu, respectively. All observation errors are assumed to be uncorrelated.

**Data Records**

The 30-year (1993-2022) NAOR is stored on a customized Thematic Real-time Environmental Distributed Data Services (THREDDS) server hosted by the SouthEast Coastal Ocean Observing Regional Association (SECOORA: https://thredds.secoora.org/thredds/catalog/secoora-cnaps-30-year/catalog.html). The data record, in NetCDF format, is divided into 30 subset files, each representing an individual year. Key OR variables in each yearly file include:

- Two-dimensional ocean bathymetry (h).

- Three-dimensional variables: a. sea surface height (zeta), b. depth-averaged zonal component of ocean velocity (ubar), c. depth-averaged meridional component of ocean velocity (vbar).
- Four-dimensional variables: a. ocean salinity (salt), b. ocean potential temperature (temp), c. zonal ocean velocity component (u), d. meridional ocean velocity component (v).

The data structure follows the format (for example, year 1993):

```
Dataset {
   Float64 h[eta_rho = 840][xi_rho = 1060];
   Float64 lat_rho[eta_rho = 840][xi_rho = 1060];
   Float64 lat_u[eta_u = 840][xi_u = 1059];
   Float64 lat_v[eta_v = 839][xi_v = 1060];
   Float64 lon_rho[eta_rho = 840][xi_rho = 1060];
   Float64 lon_u[eta_u = 840][xi_u = 1059];
   Float64 lon_v[eta_v = 839][xi_v = 1060];
   Float64 ocean_time[ocean_time = 365];
   Float32 salt[ocean_time = 365][s_rho = 50][eta_rho = 840][xi_rho = 1060];
   Float32 temp[ocean_time = 365][s_rho = 50][eta_rho = 840][xi_rho = 1060];
   Float32 u[ocean_time = 365][s_rho = 50][eta_u = 840][xi_u = 1059];
   Float32 ubar[ocean_time = 365][eta_u = 840][xi_u = 1059];
   Float32 v[ocean_time = 365][s_rho = 50][eta_v = 839][xi_v = 1060];
   Float32 vbar[ocean_time = 365][eta_v = 839][xi_v = 1060];
   Float32 zeta[ocean_time = 365][eta_rho = 840][xi_rho = 1060];
} secoora-cnaps-30-year/NCSU_CNAPS_1993.nc;
```

To read data from an online THREDDS server, several programming languages and environments may be used. Common approaches include:
- Python:
    o Use the xarray library with netCDF4 or h5netcdf backend.
    o The siphon library, specifically designed for accessing THREDDS servers.
- R:
    o Use the ncdf4 package.
    o The thredds package, designed for working with THREDDS servers.

- MATLAB:
  - Use built-in NetCDF functions.
  - The nctoolbox toolkit for more advanced operations.
- Julia:
  - Use the NetCDF.jl package.

# Technical Validation

**Innovation Errors**

Innovation errors refer to the differences between assimilated observations and the corresponding model results. No systematic trend in bias or mean absolute deviation is evident in the model innovation errors for any of the observed variables over the 30-year period (Figure 3). However, spikes in mean absolute deviation (MAD) highlight the forecast deviations that correspond to extreme events, such as tropical cyclones, that are challenging to reproduce for model and atmospheric forcing. Other sources of error include inaccuracies in the scale or timing of synoptic weather patterns, as well as errors associated with the model and/or observations.

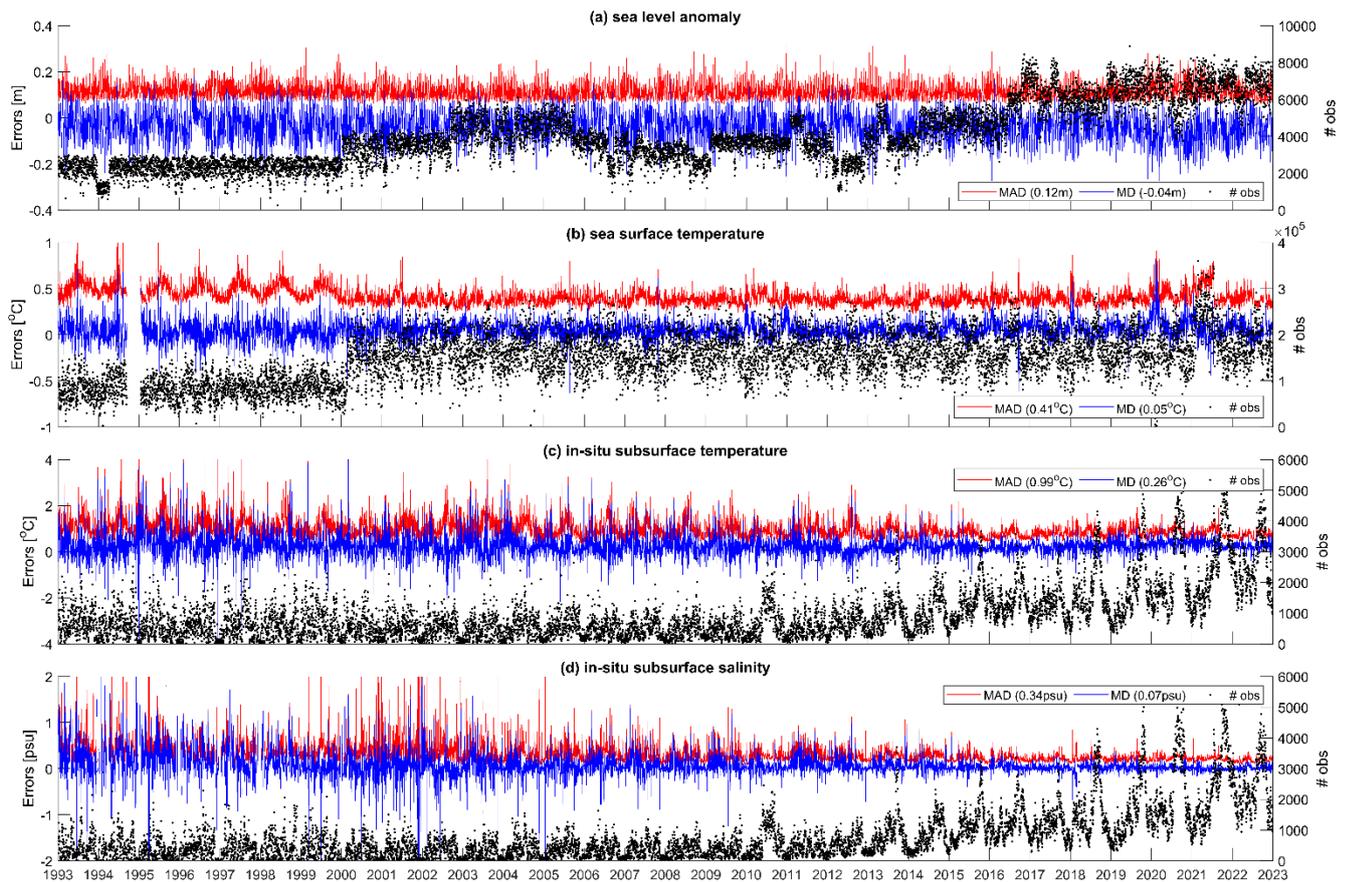

Figure 3. Time series of innovation errors for 1993–2022 and number of observations per 3-day cycle for SLA (a), SST (b), in-situ subsurface temperature (c), and salinity (d). Errors for in-situ subsurface temperature and salinity are for all depths. Mean Absolute Deviation (MAD) is shown in red, Mean Deviation (MD) is shown in blue, and number of observations per cycle (# obs) is shown in black. Mean values of MAD and MD are displayed in the legends.

**Validations Against Independent Observations**

The NAOR has been validated against various observational datasets to assess its skill in reproducing realistic ocean dynamics. Here we present a suite of validation results against independent observations that were not integrated into the reanalysis through DA.

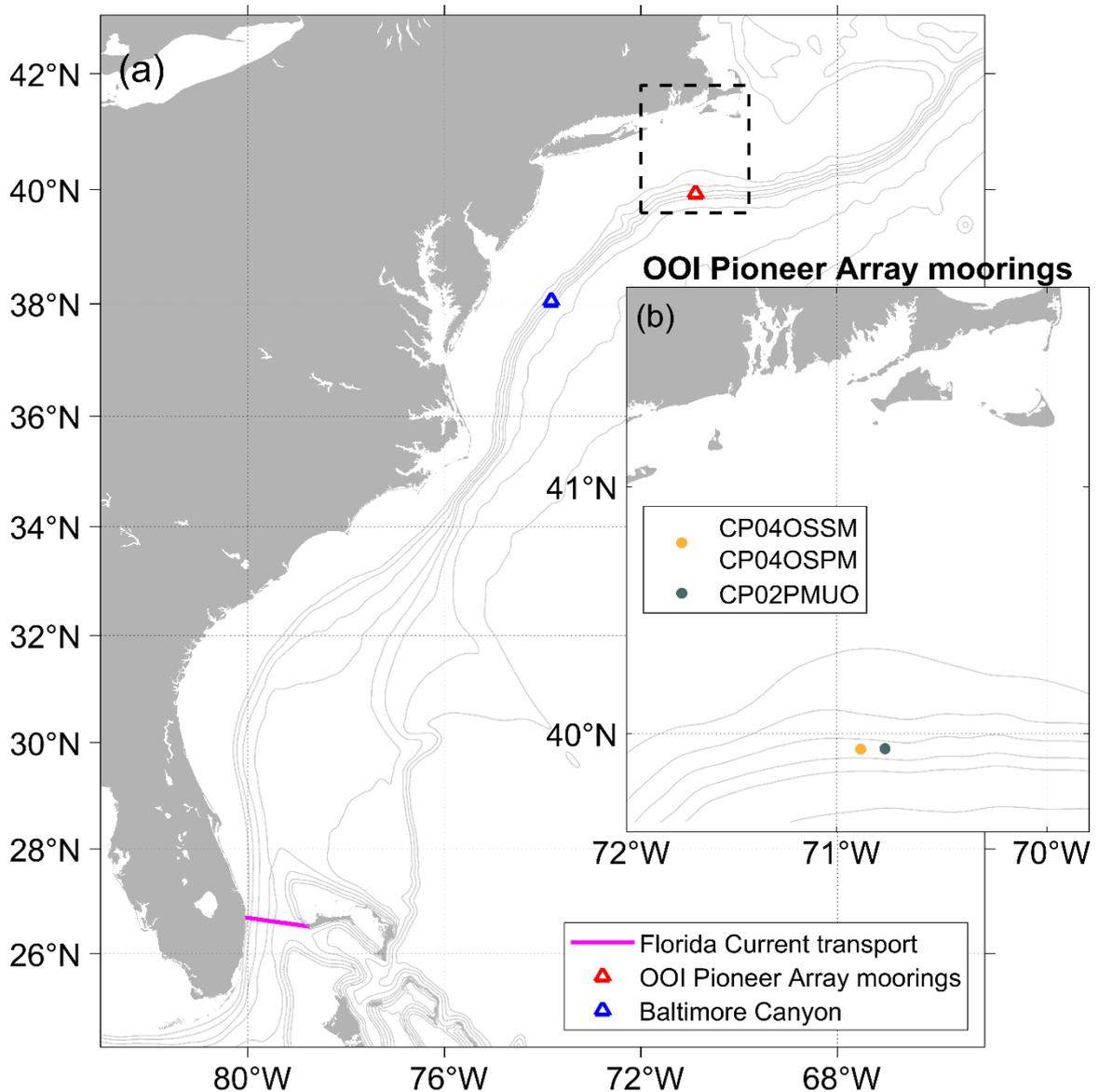

Figure 4. Observation locations of the independent timeseries validations. (a) Magenta line represents the cable measurement of Florida Current transport. The blue triangle marks the Baltimore Canyon bottom temperature observation. The red triangle represents the general location of three offshore moorings in the Ocean Observatories Initiative (OOI) Pioneer Array, the details of which are shown in (b). Gray contours represent isobaths. (b) Enlarged view of the dashed area in (a) showing the locations of OOI Pioneer Array moorings. The yellow dot represents both the offshore mooring (CP04OSPM) and the offshore surface mooring (CP04OSSM). They are 54 meters apart. The dark blue dot represents the upstream offshore mooring (CP02PMUO).

Bottom temperature observations were collected at two sites in the Mid-Atlantic Bight. First, the Ocean Observatories Initiative (OOI) Pioneer Array bottom temperature time series was collected by a CTD in the Seafloor Multi-Function Node of the offshore surface mooring (CP04OSSM). This mooring was part of the OOI Coastal Pioneer New England Shelf Array, anchored at 39.9371°N, 70.887°W, with the instrument located near the seafloor at 450 m depth (Figure 4). Quality-controlled daily averaged temperatures from December 11, 2014, to April 10, 2022, were retrieved from https://dataexplorer.oceanobservatories.org/. Second, the Baltimore Canyon bottom temperature time series was collected by a tilt current meter (TN-391) placed on the seafloor, located at 38.0481°N, 73.8226°W, and 389 m depth (Figure 4). The time series comprises quality-controlled daily averaged temperatures from March 9, 2020, to October 27, 2022. It was part of the Seep Animal Larval Transport project funded by the National Science Foundation grant OCE 185421 (https://sites.google.com/ncsu.edu/salt).

Corresponding NAOR bottom temperatures were interpolated from the daily temperature fields of the reanalysis. At both sites, the simulated bottom temperatures demonstrated very low model bias (~0.07°C) and good correlation coefficients (>0.6) against the observations (Figure 5). Notably, the reanalysis successfully captured the timing and magnitudes of warming spikes observed at the seafloor, enabling further research on bottom marine heatwaves [29]. Daily bottom temperatures at both sites were also interpolated from GLORYS and compared with NAOR (Supplementary Figure 1-2). NAOR has shown significant improvements in reproducing bottom temperature variability at both sites.

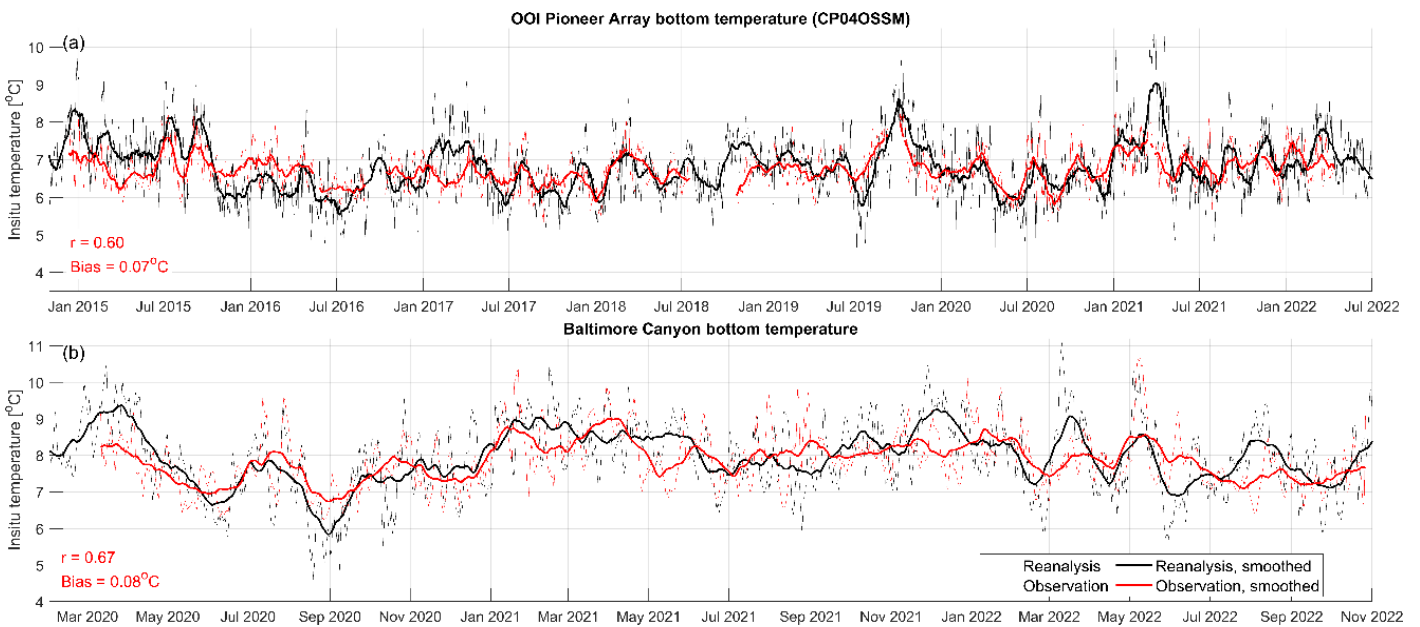

Figure 5. NAOR validation against observed bottom temperature time series at the OOI Pioneer Array and Baltimore Canyon sites. (a) Comparison of modeled and observed bottom temperatures at the OOI Pioneer Array mooring (CP04OSSM). Thin red and black lines represent daily bottom temperatures from observations and reanalysis, respectively. Bold red and black lines represent the 30-day moving averages of the corresponding temperature time series. The correlation coefficient and bias (reanalysis minus observation) between 30-day moving average time series are displayed in the bottom left. (b) Same as (a) but for the Baltimore Canyon site.

A good model representation of ocean dynamics at the continental slope is crucial for understanding shelf-ocean exchange processes. Temperature profiles at the continental slope off the New England Shelf were collected by wire-following profiler CTDs at two OOI Pioneer Array offshore moorings: CP04OSPM and CP02PMUO (Figure 4). Comparisons between NAOR and observed temperature profiles reveal that NAOR has good skill in reproducing temperature variability at all depth layers (Supplementary Figure 3-4). Throughout the water column, NAOR demonstrates lower root mean square errors (from 1.5°C at surface to 0.5°C at 300 m) and better correlation coefficients (from 0.9 at surface to 0.6 at 400 m) against the observations at both sites. GLORYS has similar performance as NAOR at the surface, likely due to the constraints from assimilated SST, while the skill decreased substantially with depth (Supplementary Figure 3-4).

The Gulf Stream, the strongest ocean current in the Northwestern Atlantic, plays a key role in modulating regional ocean dynamics. However, numerical models face challenges in accurately reproducing the Gulf Stream, particularly its separation point near Cape Hatteras and subsequent flow pattern [30,31]. Proper representation typically requires fine model resolution and realistic model bathymetry [31,32]. We evaluated the long-term mean surface current speed from our reanalysis by comparison to drifter-derived climatological surface current speed [33] (Figure 6), distributed by the Atlantic Oceanographic and Meteorological Laboratory of the National Oceanic and Atmospheric Administration (NOAA/AOML). Owing to its high model resolution, realistic bathymetry, and data assimilation, our OR reanalysis successfully reproduced the Gulf Stream separation and path after Cape Hatteras. It enables our team to conduct further research on the prediction of the Gulf Stream path using a data-driven approach [34,35].

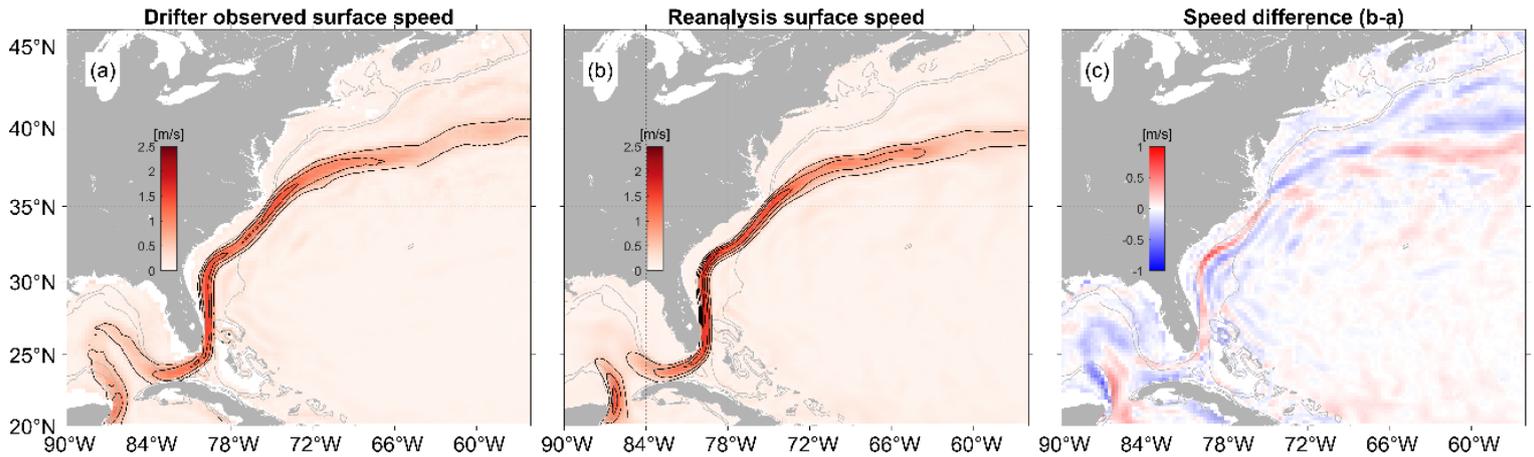

Figure 6. NAOR validation of near-surface current speed. (a) Drifter-observed climatology of near-surface current speed. (b) Mean surface current speed from the NAOR. (c) Speed difference (reanalysis minus observations).

The Western Boundary Time Series Project and the Oleander Project have provided crucial long-term routine transect observations of the Gulf Stream for decades. The Florida Current flows northward through the Straits of Florida and feeds the Gulf Stream. Underwater cable measurements from the Western Boundary Time Series Project by NOAA/AOML have provided daily Florida Current transport near 27°N since 1982 (Figure 4; https://www.aoml.noaa.gov/phod/floridacurrent/index.php). A comparison of monthly average Florida Current transport between NAOR and cable measurements (Figure 7) shows that the NAOR well captured the magnitude and variability of the Florida Current transport from 1993 to 2022, with a good correlation coefficient (>0.6) and a small root mean squared error (2.84 Sv). A comparison of daily Florida Current transport between NAOR and GLORYS demonstrates that NAOR has better skill in reproducing Gulf Stream transport variability (Supplementary Figure 5).

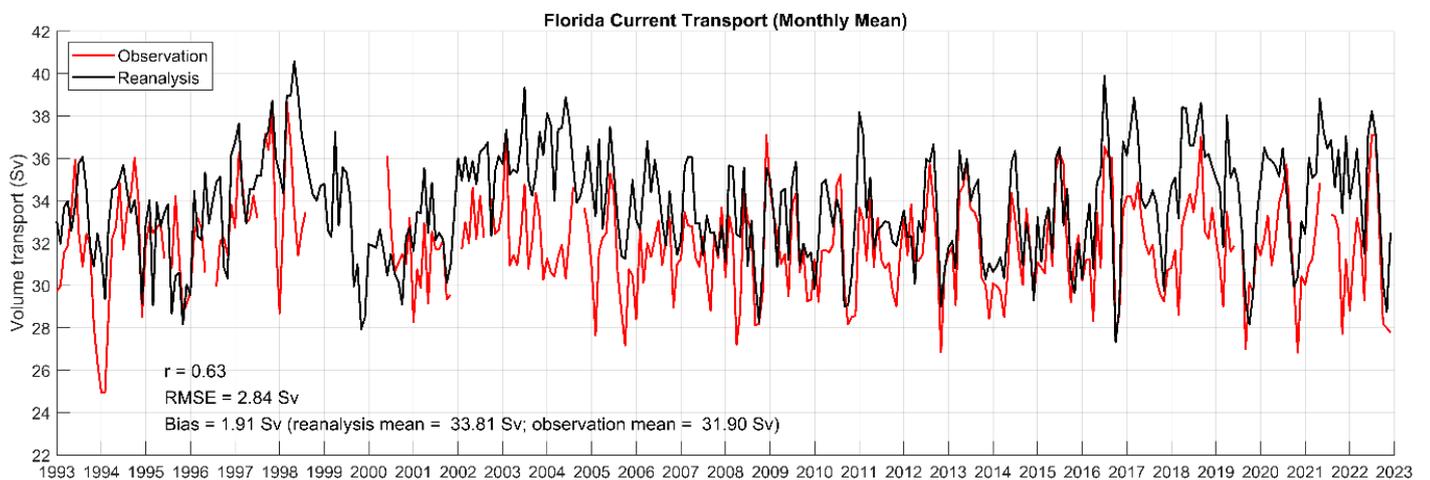

Figure 7. NAOR compared to Florida Current transport observations. Red and black lines represent monthly mean Florida Current transport from underwater cable measurements and from the reanalysis, respectively. Correlation coefficient (r), root mean squared error (RMSE), and model bias are displayed at the bottom left.

The Oleander Project has provided sustained observations along a commercial shipping route between New York and Bermuda since 1992 (http://po.msrc.sunysb.edu/Oleander). This transect crosses the Slope Sea, the Gulf Stream, and a portion of the Sargasso Sea (Figure 8a). Upper ocean velocities were routinely observed by an acoustic Doppler current profiler (ADCP) aboard a container vessel, *MV Oleander*. A gridded upper ocean velocity dataset from the project was used in the validation [36] (https://zenodo.org/records/3935983). The dataset contains gridded ADCP current velocity from 1994-2018 with a 4 km horizontal resolution and an 8 m vertical resolution. Comparison of the long-term averaged upper ocean velocity between the reanalysis and the Oleander Project observations (Figure 8) shows that the reanalysis accurately reproduced the location, speed, and direction of the Gulf Stream and the Shelfbreak Jet, confirming the model's skill in representing the Slope Sea circulation and Gulf Stream structures.

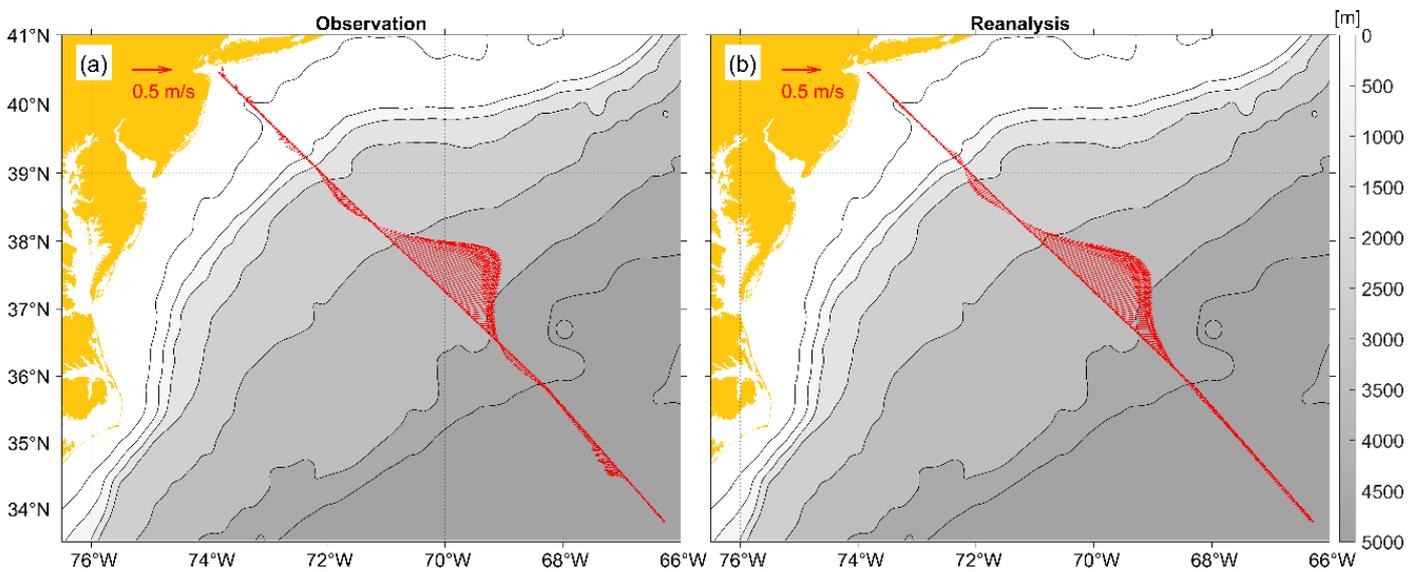

Figure 8. NAOR compared to the upper ocean velocity observations from the Oleander Project. (a) Red arrows represent the observed climatological upper ocean velocity averaged over the depth between 26 m and 202 m. (b) Same as (a) but for the reanalysis.

Similar current velocity validations were also performed against mooring data collected from the Gulf of Mexico [37–39]. The Loop Current enters Gulf of Mexico through the narrow

Yucatan Channel causing irregular meandering and eddy shedding events which impact the Gulf [40–42]. The comparison between modeled and observed subsurface mean velocity vectors and variance ellipses between 50 m and 3300 m (Figure 9) shows that, despite overestimation of mean current across the Yucatan Channel, the NAOR reproduced well the variability and vertical structure of the Loop Current both at the Yucatan Channel and in the Gulf of Mexico. Future works are underway to improve the Yucatan Current by acquiring and incorporating additional observations and fine-tuning model parameters. Using this long-term NAOR dataset, our team also developed a machine learning model capable of predicting Loop Current and Loop Current eddy shedding over a long lead time [34,43]. The NAOR also facilitates interdisciplinary studies in regions with complex coastline and bathymetry, such as the recruitment of key reef species in the Florida Keys [44].

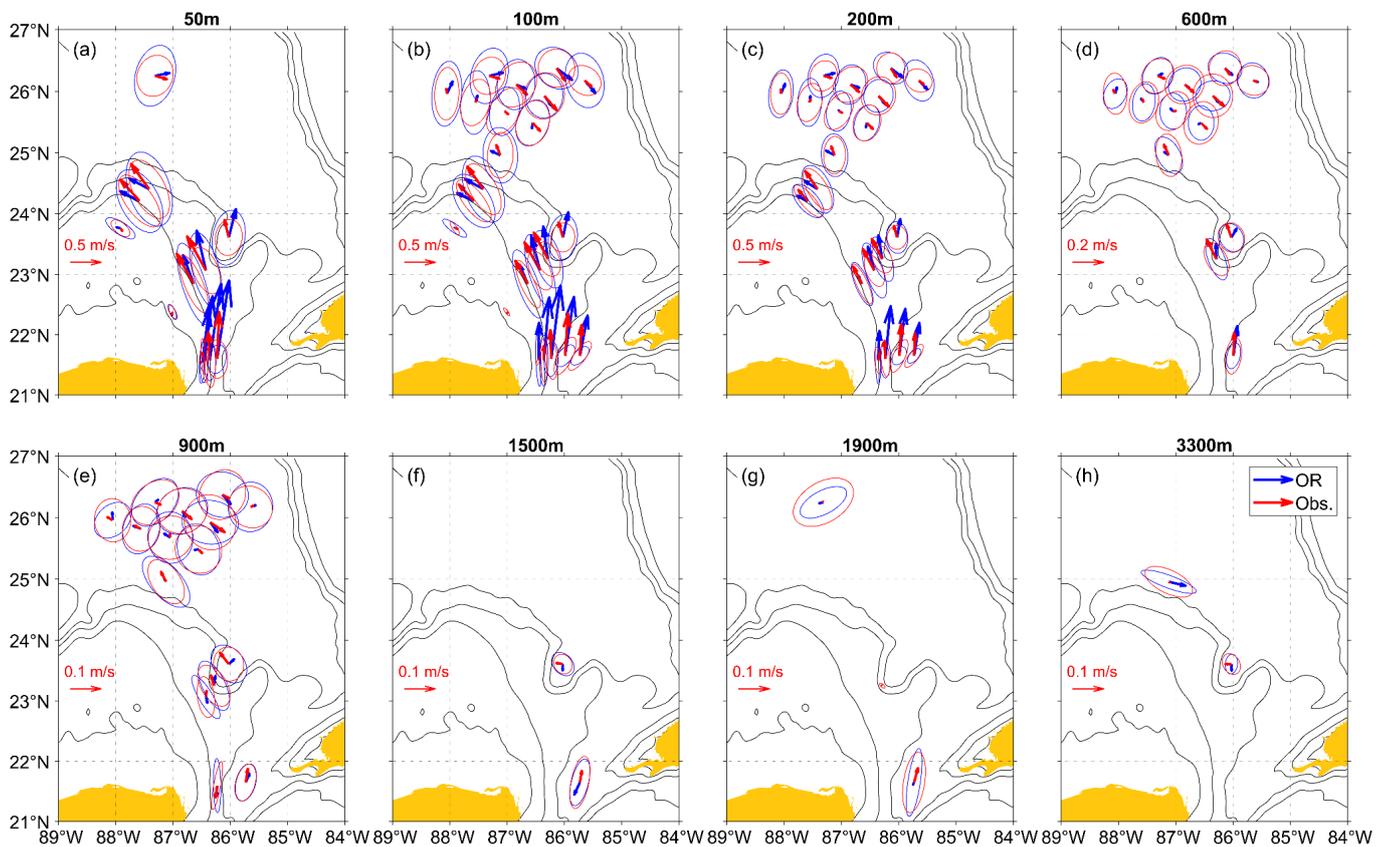

Figure 9. NAOR validation against ocean current observations in the eastern Gulf of Mexico at depths from 50 m (a) to 3300 m (h). At each station, mean velocity vectors and variance ellipses are shown for current meters (red) and NAOR (blue).

## Data Usage Notes

The NAOR data are available under a Data Use Agreement (DUA) that outlines users' information, scope of distribution, and notification for use or publications.

## Code Availability

The modeling system used to generate the reanalysis was ROMS version 3.9 (https://github.com/myroms/roms).

## Acknowledgements

Research support for this project provided through NSF grants OCE-1559178 and OCE-1851421 and NOAA grant NA16NOS0120028 is much appreciated. We thank data providers NASA, NOAA IOOS, CMEMS, ECMWF, and UK Met Office for providing data for modeling and assimilation, and NSF OOI Pioneer, NOAA AOML, and the Oleander Project for providing observations for validation. We also thank Dr. Julio Sheinbaum of CICESE, Mexico, for providing the Gulf of Mexico velocity means and variance data.

## Author contributions

**RH**: Conceptualization, methodology, writing, supervision, project administration, and funding acquisition. **TW**: Methodology, software, validation, formal analysis, investigation, writing, and visualization. **RH** and **TW** have contributed equally in leading and conducting this research. **SM, HZ, JB, JW**: software, data curation, and editing. **JD, DH**: Funding acquisition, data curation and archiving, and editing.

## Competing interests

The authors declare no competing interests.